\documentclass{emulateapj}

\usepackage{ulem} 
\usepackage{xcolor}
\usepackage{graphicx}
\usepackage{enumerate}

\newcommand{\crhydro}{{\it CR-hydro-NEI}}

\newcommand\Msun{M_{\odot}}

\newcommand{\be}{\begin{eqnarray}}
\newcommand{\ee}{\end{eqnarray}}

\shorttitle{Progenitor Models for Core--Collapse Supernova Remnants}

\shortauthors{Patnaude et al.}

\begin{document}

\title{Are Models for Core--Collapse Supernova Progenitors Consistent
with the Properties of Supernova Remnants?}

\author{Daniel J.~Patnaude\altaffilmark{1}, 
Shiu-Hang Lee\altaffilmark{2}, Patrick O.~Slane\altaffilmark{1},
Carles Badenes\altaffilmark{3},
Alexander Heger\altaffilmark{4}, Donald C.~Ellison\altaffilmark{5} and
Shigehiro Nagataki\altaffilmark{6}}
\altaffiltext{1}{Smithsonian Astrophysical Observatory, Cambridge, 
MA 02138, USA; dpatnaude@cfa.harvard.edu; pslane@cfa.harvard.edu}

\altaffiltext{2}{Institute of Space and Astronautical Science, Japan Aerospace Exploration Agency,
3-1-1 Yoshinodai, Chuo-ku, Sagamihara, Kanagawa 252-5210, Japan; 
slee@astro.isas.jaxa.jp}

\altaffiltext{3}{Department of Physics and Astronomy and Pittsburgh 
Particle Physics, Astrophysics and Cosmology Center (PITT PACC), 
University of Pittsburgh, 
3941 O'Hara St, Pittsburgh, PA 15260, USA; badenes@pitt.edu}

\altaffiltext{4}{Monash Center for Astrophysics, School of Mathematical 
Sciences, Bldg 28, Monash University, Vic 3800, AU; alexander.heger@monash.edu}

\altaffiltext{5}{Physics Department, North Carolina State
University, Box 8202, Raleigh, NC 27695, USA;
don\_ellison@ncsu.edu}

\altaffiltext{6}{RIKEN, Astrophysical Big Bang Laboratory, 
2-1 Hirosawa, Wako, Saitama 351-0198, Japan; shigehiro.nagataki@riken.jp}

\begin{abstract}

The recent discovery that the Fe-K line luminosities and energy centroids
observed in nearby SNRs
are a strong discriminant of both progenitor type and circumstellar 
environment has implications for our understanding of supernova progenitor 
evolution. 
Using models for the chemical composition of core--collapse supernova ejecta,
we model the dynamics and thermal X-ray emission from shocked 
ejecta and circumstellar material, modeled as an $r^{-2}$ wind,
to ages of 3000 years. We compare the X-ray spectra expected 
from these models to observations made with the {\it Suzaku} satellite.
We also model the dynamics and X-ray emission from Type Ia progenitor
models. We find a clear distinction in Fe-K line energy centroid
between core--collapse and Type Ia models. The core--collapse supernova
models predict higher Fe-K line centroid energies than the Type Ia models, 
in agreement with observations.
We argue that the higher line centroids are a consequence of the
increased densities found in the 
circumstellar environment created by the expansion 
of the slow-moving wind from the massive progenitors.
%For the isotropic winds considered here, the computed luminosities
%and energy centroids from the models are in good agreement with
%observations, but there is not good agreement between the model 
%remnant radii and observations.

\end{abstract}


\section{Introduction}

Supernovae, both thermonuclear (Ia) and core--collapse, represent
the endpoints in the stellar evolution of either Chandrasekhar white
dwarfs or massive stars with zero age main sequence masses $>$ 8M$_{\sun}$. 
Core--collapse supernovae (CCSNe) exhibit a wide diversity of properties. 
Classifications are based on spectra and light curves, and not necessarily
on a physical mechanism \citep{filippenko97}. Types I and II are determined
by the presence (II) or absence (I) of H lines in the spectrum. Type I
CCSNe include Type Ib/c, Ibn, and Ic-BL. The progenitors of these supernovae
are thought to be either massive He stars or some subset of Wolf-Rayet (WR)
\citep[see][for a broad classification]{galyam07}.
Type II SNe include the subtypes IIP \citep[e.g., 
SN~1999em;][]{smartt02} IIL 
\citep[e.g., SN~1979C;][]{fesen93}, II-pec 
\citep[e.g., SN~1987A;][]{chevalier89}, 
IIn \citep[e.g., SN~1995N;][]{fransson02}, IIb 
\citep[e.g., SN~1993J;][]{fransson96}, 
and possibly some 
superluminous supernovae that exhibit evidence for a strong circumstellar 
interaction, such as SN~2006tf \citep{smith08}.

Unlike Type I CCSNe, the progenitors of Type II supernovae
are likely red or yellow supergiants (R/YSG), or in the case of IIn, luminous
blue variables \citep[LBVs;][]{galyam07}. In at least a few
cases, progenitors have been identified in pre-explosion images 
\citep{smartt09}. 
In both Type I and II CCSNe, the progenitor
mass--loss rate can vary between 10$^{-6}$ -- 10$^{-4}$ M$_{\sun}$ 
yr$^{-1}$ (and for IIns and SLSNe, 0.1--10 M$_{\sun}$ yr$^{-1}$, 
presumably as eruptive mass--loss),
but the wind velocity ($v_w$) between the RSG or WR progenitor can differ by as
much as two orders of magnitude \citep[$v_{w}$ = 10--20 km s$^{-1}$ in
a RSG, and $\sim$ 1000 km s$^{-1}$ in a WR star;][]{crowther07,smith14}. 
For steady, isotropic
mass--loss, the circumstellar medium (CSM) density is proportional to
$\dot{M}$/$v_w$. Thus high mass--loss rates coupled with
slow wind velocities can lead to a substantial amount of mass close
to the supernova progenitor. This would be in contrast to the large,
evacuated cavities expected around WR progenitors, due to their large
mechanical wind luminosities \citep[c.f.,][]{koo92}.

While typing supernovae based on their optical spectra is well established
\citep{filippenko97},
connecting supernovae types to supernova remnants (SNRs) remains difficult.
In the case of Type Ia supernovae, their remnants can frequently be 
readily identified by the iron content observed in the X-ray spectrum
\citep{hughes95,badenes06,badenes08,patnaude12}. 
Given the broad diversity of CCSN types, connecting
remnants to supernovae (and to supernova models) is a challenge. However,
examples exist where supernovae have been linked to supernova remnants, 
or supernova remnants have been directly typed. For instance,
light echo analysis allows for direct typing of some supernova remnants
such as Cas A
\citep{fransson96,krause08,rest08}, and  
\citet{milisavljevic12} recently showed that the [\ion{O}{3}] line shape in
SN~1993J was remarkably
similar to that of Cas A. Finally, similarities in
X-ray spectra provide evidence that SN~1996cr may be a Type II~pec,
similar to SN~1987A \citep{bauer08}. However, beyond these examples,
connections between SNe and SNRs remain sparse, particularly for
the class of core--collapse SNe \citep{hughes95}.

Recently, \citet{yamaguchi14} presented a method of typing supernova remnants
based on the Fe-K$\alpha$ line centroid and luminosity. 
Since Fe is produced in the center of the progenitor
during the explosion, heating of Fe can be delayed, resulting in an
ionization state lower than He-like (Fe$^{24+}$) in young and middle-aged
SNRs. The ionization state affects the Fe-K line centroid, which can
be measured to high precision with current X-ray satellites. 
\citet{yamaguchi14} showed
that the Fe-K line centroids for Ia SNRs are generally lower ($<$ 6550 eV)
than those found in core--collapse SNRs. Additionally, they found that
when computing synthetic Fe-K line centroids and luminosities from
well tested models for Type Ia ejecta, the models 
\citep{badenes06,badenes08} predicted bulk properties
in line with observations \citep{yamaguchi14}.

Here we extend the work of \citet{yamaguchi14} to address the question
of whether models for CCSNe are able to reproduce, in broad terms, the
observable bulk properties of core--collapse SNRs. We employ a model which
tracks the hydrodynamics and time--dependent ionization of shocked 
circumstellar material and supernova ejecta, coupled to an emissivity
code to compute Fe-K line centroids and luminosity, as a function
of supernova ejecta model, age, and circumstellar environment. We 
compare our models to {\it Suzaku} observations of Galactic and
Magellanic Cloud SNRs, and discuss implications and future directions
which this work can take. 

\section{Hydrodynamical Model}

We employ our \crhydro\ code which allows us to simultaneously 
describe the thermal and non-thermal emission at the forward and 
reverse shocks in young SNRs. \crhydro\ is a 1D Lagrangian hydrodynamics
code based on VH-1, a multidimensional hydrodynamics code developed by
J.~Blondin and colleagues \citep[e.g.,][]{blondin93}. \crhydro\ simultaneously
models the supernova blastwave dynamics and particle acceleration, including
the back reaction of the nonthermal particles on the SNR dynamics. It
is capable of modeling the nonthermal particle spectrum as well as the
broadband nonthermal and thermal emission from shocked circumstellar material
and ejecta. Here, we run the model without considering
the effects of diffusive shock acceleration on the dynamics and
emitted spectra (i.e., we set the particle injection to the 
test particle limit). Specific details concerning \crhydro\ can be found in 
\citet{ellison07,patnaude09,
ellison10,patnaude10}; and \citet{lee14}. Recent modifications to the code 
allow
us to track the evolution of the thermal emission from the shocked
ejecta, as well as employ custom ejecta models \citep{lee14}. 

\subsection{Ejecta Models}

For this study, we make use of previously available models for
the composition of the ejecta in both core--collapse (CC) and
Type Ia SNe. For the Ia models, we employ the delayed detonation 
models DDTa and DDTg \citep{badenes03}. In these models, a 
flame propagates as a slow deflagration, with a transition to a
detonation induced at some flame density. The transition point sets
the amount of $^{56}$Ni produced in the explosion. The DDTa model produces 
$\sim$ 1 M$_{\sun}$ of $^{56}$Ni, while the less energetic DDTg model
produces only $\sim$ 0.3 M$_{\sun}$ of $^{56}$Ni.

To model the core--collapse supernovae, we employ a range of
ejecta models from a variety of single star progenitor scenarios. 
For this paper we have computed new stellar models of $12\,\Msun$ and
$25\,\Msun$ initial mass, Models s12D and s25D.  These models have
been computed using the KEPLER stellar evolution code \citep{weaver78}
and are similar to
those in \citet{woosley07} (see also \citealt{woosley02,rauscher02}), 
but use the updated solar
abundances from \citet{lodders09}.
For both models, the stellar evolution, including mass--loss, is
followed to the point of core--collapse.  Model s12D loses
approximately 3 M$_{\sun}$ of material, while Model s25D loses
$\approx$ 8 M$_{\sun}$ due to stellar winds by the time of 
core--collapse.  Details of the explosion can be found in
\citet{rauscher02}.  As a brief summary, a piston is placed a
the base of the oxygen burning layer where entropy per baryon rises
above $S/k_B=4$.  The piston is first moved inward at about $1/4$ of
local gravitational acceleration until a radius of 100 km is
reached; then it is moved outward at a constant fraction of local
acceleration until a radius of 10,000 km is reached.  The outward
acceleration is adjusted such that an explosion energy of
1.2$\times$10$^{51}$ erg is reached; the estimated explosion energy
for SN 1987A is taken as a typical value here.  The prescription for
mixing in the ejecta can be found in \citet{heger10}, 
but the default parameters used here have been adjusted such
that enough $^{56}$Ni is mixed into the envelope to reproduce the light
curves of Type II supernovae, in good agreement with hydrodynamic
simulations \citep{joggerst09}.

We also use a model for CCSN ejecta tailored to SN~1987A 
\citep{saio88,hashimoto89,shigeyama90}. The model comprises a 
6 M$_{\sun}$ Helium star
enclosed in a 10 M$_{\sun}$ hydrogen envelope. Finally, we include a
model that has been previously applied to the Type IIb SN~1993J 
\citep{shigeyama94}. The progenitor for this model is an 18 M$_{\sun}$ main sequence
progenitor with a metallicity $Z = 0.02 Z_{\sun}$. This progenitor loses
$\approx$ 15 M$_{\sun}$ of material prior to core--collapse. Further details of
 the models are summarized in Table~\ref{tab:models}.

Though
\citet{yamaguchi14} have previously shown agreement between Type Ia
models and measured Fe-K line centroids and luminosities, we include
a subset of Ia models in our sample to validate our results. For reasons
discussed below, we do not expect an exact agreement between our Ia 
results and those in \citet{yamaguchi14}, but broad agreement is provided
as a consistency check. The core--collapse SN ejecta models are from
the evolution of single-star progenitors and thus do not include
the effects that binarity have on the progenitor evolution and
the circumstellar environment \citep{sana12}. We will address the 
role of binarity on both the progenitor and CSM in a future paper.

\subsection{Circumstellar Environments}

A key question is the shape of the circumstellar environment around
the supernova progenitor. Red and yellow supergiants (R/YSG) expel
several solar masses of material over their lifetimes (10$^{-6}$ -- 10$^{-4}$ 
M$_{\sun}$ yr$^{-1}$) with velocities of 10--100 km s$^{-1}$. 
Wolf-Rayet progenitors expel similar 
amounts of mass per year, but with wind velocities 100$\times$ that
of their R/YSG counterparts \citep[e.g.,][]{smith14}. Additionally,
episodic mass--loss may be relevant in the thermal and dynamical 
evolution of SNe \citep{pan13}, resulting in emission that is luminous
early on, but decays sharply due to a sudden decrease in the circumstellar
density \citep[e.g.,][]{dwarkadas11,dwarkadas12}. Regardless of
details, it is currently accepted that SNe of the Type II variety
expel material with velocities $\sim$ 10 -- a few hundred km s$^{-1}$,
while core--collapse of the Ib/c (and variants) have wind velocities
$\sim$ 1000 km s$^{-1}$ \citep{smith14}.

Given the diverse nature of the CSM environment expected around 
CCSNe progenitors, we chose to take a general approach and consider a 
range of mass--loss rates and wind velocities appropriate for massive, 
red supergiant progenitors. We consider
progenitor winds with mass--loss rates of $\dot{M}$ = 1 -- 2$\times$10$^{-5}$
M$_{\sun}$ yr$^{-1}$, and wind velocities of $v_w$ = 10 -- 20 km s$^{-1}$. 
We do not consider CSM environments
shaped by fast winds, {\it e.g.,} Wolf-Rayet and O-type stars, as these
winds are expected to result in low density cavities 
\citep[see, e.g.,][]{dwarkadas07}. 
The low density CSM does not lend itself to 
substantial thermal X-ray emission 
\citep[e.g., as in SNR RX J1713.7-3946][]{ellison12}, 
and any shocked CSM has a low
temperature due to the long Coulomb heating timescale.
For the Ia models, we assume uniform 
densities with ranges of 0.3 -- 3.0 cm$^{-3}$, consistent
with results inferred from observations of the environments around
Ia progenitors \citep{badenes07}.

\section{Modeling and Results}

We evolve the CC SNe to ages of 3000 years. \citet{dwarkadas98}
pointed out that for a Type Ia ejecta model interacting with a constant
density ambient medium, the reverse shock propagates to the center
of the ejecta when the forward shock has swept up $\sim$ 24$\times$
the ejecta mass. For the Ia models considered here, the reverse shock
reaches the ejecta center at ages of $\sim$ 1000 years for the
densest CSM environments. Thus we only consider the evolution of 
the Type Ia models to 1000 years. This differs from \citet{yamaguchi14}
where they evolve the Ia models to much larger ages, allowing the shock
to bounce and re-shock previously shocked ejecta. For both the
CC and Ia models, we evolve them over the range of parameters summarized
in Table~\ref{tab:models}, resulting in a grid of models that span
a range of circumstellar environments.

The ejecta models include not only the composition as a function
of mass coordinate, but also the density and velocity as a function
of mass coordinate. For this study, we have mapped the ejecta composition
onto approximations of the ejecta structure: the core--collapse 
supernova remnant models assume ejecta with a constant density 
core and an $n=9$ power law envelope \citep{truelove99}. The
Type Ia ejecta are modeled with an exponential ejecta profile 
\citep{dwarkadas98}. 
Our code produces ionization fractions for both shocked CSM and ejecta
as a function of SNR age. We pass these parameters to an emissivity code
\citep{patnaude10,lee14}
to compute the full SNR spectrum as a function of time. As discussed 
in \citet{lee14},
we include thermal and Doppler broadening, though we do not consider
their effects on our results here, nor do we consider the individual
contributions from shocked CSM versus shocked ejecta as a function of time.
Future X-ray observatories such as {\it Astro-H} may be able to 
discern between these two components. At chosen time-steps, we used
{\texttt Xspec}\footnote{http://heasarc.gsfc.nasa.gov/xanadu/xspec/} 
to synthesize a {\it Chandra} ACIS-S observation of the
SNR. We then fit the 6 -- 7 keV emission from the simulated observation
to a Gaussian with a power-law continuum.\footnote{Since we are fitting the
simulated spectrum over such a narrow band pass, the continuum, 
which in reality is thermal in nature and the sum of several shocked
components, is well approximated by a power-law.} The results from these
simulations, for the grid of Type Ia and core--collapse SNe 
models lists in Table~\ref{tab:models} are shown in 
Figure~\ref{fig:models}, where we plot the Fe-K line centroid as
a function of luminosity (left panel), as well as the evolution of
the Fe-K line luminosity as a function of scaled SNR age (right panel). 

\section{Discussion}

The data in Figure~\ref{fig:models} (left) are taken from
\citet{yamaguchi14}. In broad terms, we find results that are consistent
to those in \citet{yamaguchi14}: supernova remnants typed by their Fe content
as Ias are generally consistent with the synthesized X-ray spectra from
Ia models. Note that we do not expect an exact match between the 
results presented in \citet{yamaguchi14} and here: our calculations include
emission from both shocked ejecta and circumstellar material, while
they only consider emission from shocked ejecta; the electron to proton
temperature at the shock in their model is assumed to be either 0.01 or
0.03, while we assume mass proportional heating at the shock, followed
by Coulomb heating downstream; as discussed in \citep{patnaude10},
the collisional ionization and recombination rates used in the ionization 
balance calculations, as well as the atomic data used in the spectral
synthesis likely also differ. 
However, even with these numerous differences, it is encouraging to 
find agreement between our Ia results and those presented in 
\citet{yamaguchi14}, and gives us confidence that the results for
the core--collapse models are insensitive to these differences. 
The line centroids and luminosities for the core--collapse models
are also presented in Figure~\ref{fig:models} (left). For those
models, we find that the measured Fe-K luminosities
and line centroids are consistent with the models, with a few notable
exceptions discussed below. 

As discussed in \citet{yamaguchi14}, all Ia SNRs have observed Fe-K line
centroids $\lesssim$ 6550 eV, and the chosen Ia ejecta models also do
not produce Fe-K centroids in excess of 6550 eV. Even Kepler's SNR,
which is thought to be the result of a luminous Ia in a modified
CSM \citep{chiotellis12,patnaude12} still displays bulk properties
consistent with a Ia origin. 
%The reason for this is because, while
%this particular SNR may be evolving into a dense wind, the ejecta
%in Type Ia SNRs have a lower density than in CC SNRs -- collisional
%ionization and heating of the ejecta is thus less efficient.

For the observed core--collapse SNRs, the observed line centroids are 
broadly consistent with 
the model predicted centroids and luminosities. 
In most cases the core-collapse models show Fe line centroids that are in 
excess of 6550 eV. Only in the youngest models ($t_{\mathrm{SNR}}$ $\lesssim$
200 yr) do we find Fe line centroids at or below 6550 eV. 
\citet{yamaguchi14} have argued that the higher observed line centroids 
are likely the result of higher ambient medium density, and our results support
this conclusion.
For the highest density
Ia model, $n_{amb}$ = 3.0 cm$^{-3}$, while for the core--collapse 
models, even though the density follows a $r^{-2}$ power law, 
$n_{amb}$ $>$ 10 cm$^{-3}$ 
at radii in excess of 10$^{17}$ cm, owing to the high density
close in to the progenitor. The higher ambient 
medium density has a two-fold effect: first, the shocked CSM is at 
a higher temperature and ionization state, since heating $\propto$ 
$n^2$, and ionization rates are $\propto$ $n$; secondly, the
dense circumstellar environment produces a strong reverse shock
which can deposit more energy into the ejecta, resulting in
higher overall charge states in the shocked ejecta.

In Figure~\ref{fig:models} (right), we plot the time evolution of
the Fe-K emission from both the forward and reverse shocks, for the
range of mass--loss parameters listed in Table~\ref{tab:models}. We scale
the SNR age to the characteristic SNR age given by \citet{truelove99}:

\begin{equation}
t_{\mathrm{ch}} = 1770E_{51}^{-1/2}\left(\frac{M_{\mathrm{ej}}}{M_{\sun}}\right)^{3/2}\dot{M}^{-1}_{w,5}v_{w,6} \ \mathrm{yr} \ ,
\end{equation}

\noindent
where the ejecta masses and explosion energetics are taken from 
Table~\ref{tab:models}. As seen in Fig.~\ref{fig:models} (right), 
the emission from swept up circumstellar and shocked ejecta is
initially comparable in the s12D, s25D, and SN~1993J models, but
the emission from shocked ejecta quickly surpasses that of the 
shocked circumstellar material by a factor of two or more. 
Interestingly, and previously pointed out in \citet{lee14}, there
is no appreciable Fe-K emission from shocked ejecta in the SN~1987A 
models, even
at late times (our models do show that the Fe-K emission from shocked
ejecta does begin to appear at late times, though remains very weak). 
The SN~1987A model differs from the other models in two aspects: first,
it is a compact blue progenitor, and the densities in the ejecta 
are higher than in the other models with extended H-rich envelopes, and
secondly, the Fe abundance in the SN~1987A model is $\sim$ two orders
of magnitude lower than in the other models-- there is not a significant
amount of Fe to shock in the ejecta.

We also investigate the relationship between the Fe-K line luminosity
and energy centroid and the SNR radius. In Figure~\ref{fig:dyn_spec}
we plot the line luminosity as a function of radius 
for each model (left panels)
while in the right panels we plot the line energy centroid as a 
function of radius. Note that for those models with a high mass--loss
rate and low wind velocity, the Fe-K line centroid approaches the
collisional ionization equilibrium value in several hundred years
While the measured line centroids as a function
of SNR radius do generally agree with the models, the observed 
line luminosities for as many as half of the sampled remnants are greater
than the modeled luminosities by as much as an order of magnitude. 
We discuss these results in more detail below.

%We also investigate the relationship between Fe-K line centroid and
%SNR radii. Shown in Figure~\ref{fig:models} (right), we plot the
%Fe-K line centroid as a function of SNR radius for both the data
%and the CCSNe models shown in Figure~\ref{fig:models} (left). 
%With some exceptions, such as G292.0+1.8 
%(though this SNR exhibits one of the lowest line luminosities), the
%models again agree in estimating remnant radius versus 
%Fe-K energy centroid. That G292.0+1.8 lies in the middle of the
%energy versus luminosity parameter space, but lies outside of the 
%radius versus energy centroid space suggests a more complex CSM 
%interaction, as evidenced by the bright CSM ring observed in X-rays
%\citep{lee10}. However, uncertainty in the radius \citep[7.7 pc
%vs 11 pc;][]{park04,lee10} may bring the data in line with the models.

\citet{chevalier05} typed several Galactic SNRs based on the properties
of the swept up circumstellar material, classifying many as either Type IIb/L
or Type IIP. Main sequence progenitors 
for Type IIP and IIb/L SN span a mass range of 10--25 
M$_{\sun}$, and have slow dense winds resulting in much of the lost
mass remaining close to the progenitor. Outside of this is a low 
density bubble created by the fast wind from the main--sequence phase.
Models s12D and s25D represent such RSG progenitors \citep{heger10}. 
Model s25D has lost almost
13M$_{\sun}$ of material over the course of its evolution, indicating
that the RSG phase for this progenitor cannot be more than $\sim$ 10$^{6}$
yr, though it is probably less than this. The extent of the region 
occupied by the RSG wind is set by the pressure of the surrounding
interstellar medium, and can range from $\lesssim$ 1 pc in IIP 
progenitors to greater than 5 pc in IIb/L progenitors \citep{chevalier05}.
The modeled radii shown in Figure~\ref{fig:dyn_spec} are generally
consistent with this, and may suggest that at radii in excess of
10 pc, the wind from an earlier phase of evolution is required -- 
as discussed in Section~2, we applied 
these models with a broad parameter space, and more detailed modeling
for each individual object is necessary.

\subsection{Outliers}

As seen in Figure~\ref{fig:models} (left), while there is generally
good agreement between the supernova explosion models and the bulk 
properties, several core--collapse 
measurements stand out as having higher than average
Fe-K line luminosities: Cas A, N132D, W49B, and N63A. In the case
of Cas A, it appears likely that there was bulk overturning of the
deepest layers of ejecta during the explosion \citep{hughes00}. This
resulted in the Fe-rich ejecta being brought to the surface during 
the explosion. Subsequently, the iron is observed to have been shocked
at early times, resulting in a high charge state and luminosity
\citep{hwang03,hwang12}. The two LMC SNRs N132D and N63A are both thought
to be interacting with a large amount of interstellar material in
star forming regions, resulting in copious swept up shocked material
\citep{warren03,borkowski07}.

W49B also displays higher than average Fe-K luminosities, when compared
to the other CC SNRs. \citet{lopez13a} postulated that W49B is the result
of a bipolar Type Ib/c SN, based on the morphology and other spectral
characteristics. However, the low CSM densities inferred from X-ray and radio 
observations to be
around Ib/c progenitors \citep[e.g., SN~2007gr;][]{soderberg10} 
appears incongruent with a high Fe-K charge state
and line luminosity in the spectrum of W49B. 
Type Ib/c SNe probably result from Wolf-Rayet stars
with fast winds and high mechanical luminosities. Compared to red and
yellow supergiant progenitors, the 100$\times$ faster wind velocities
in WR progenitors
transport much of the circumstellar material to larger distances, leaving
behind a largely low density CSM cavity, 
though evidence does exist for more complex
environments around some SN~Ic, such as in the case of SN~2007bg, which
appears to have underwent several differing mass--loss phases 
prior to the SN \citep{salas13}. 

Additionally, \citet{ozawa09} found evidence for an overionized 
plasma in W49B, as evidenced by the ratio of H-- to He--like lines
\citep[see also][]{lopez13b,yamaguchi14}. 
\citet{moriya12} recently presented a model for 
SNRs that show overionized plasmas, and suggested that red supergiants
can deposit the mass required for rapid ionization 
close enough to the progenitor to 
overionize the shocked material at early times. He noted that
the time required to reach ionization equilibrium in 
a RSG wind is $\sim$ 100$\times$ faster than in a W-R wind, unless
collisional ionization equilibrium in the W-R wind occurs almost 
immediately after the explosion \citep{moriya12}.
However, an emergent class of supernovae
typed as Ibn has recently been discovered that displays a strong
circumstellar interaction like those found in IIn's as well as 
no H in the early time optical spectra, like Ib's, such as SN~2006cj
\citep{chugai09} and SN~2011w \citep{smith12}. The progenitors of these
systems include Wolf-Rayet and LBVs with zero age main sequence masses
M $>$ 40 M$_{\sun}$, and possibly display eruptive events
just prior to the SN \citet{smith14}. 
A progenitor such as this may explain both the overionization 
and high Fe abundance observed in W49B and
reconcile these apparently contradictory results with theory. In any event,
detailed modeling of a W49B-like progenitor, explosion, and 
subsequent evolution might be required to explain this odd object.

\subsection{Spectral vs. Dynamical Quantities}

\begin{figure}
\includegraphics[width=0.5\textwidth]{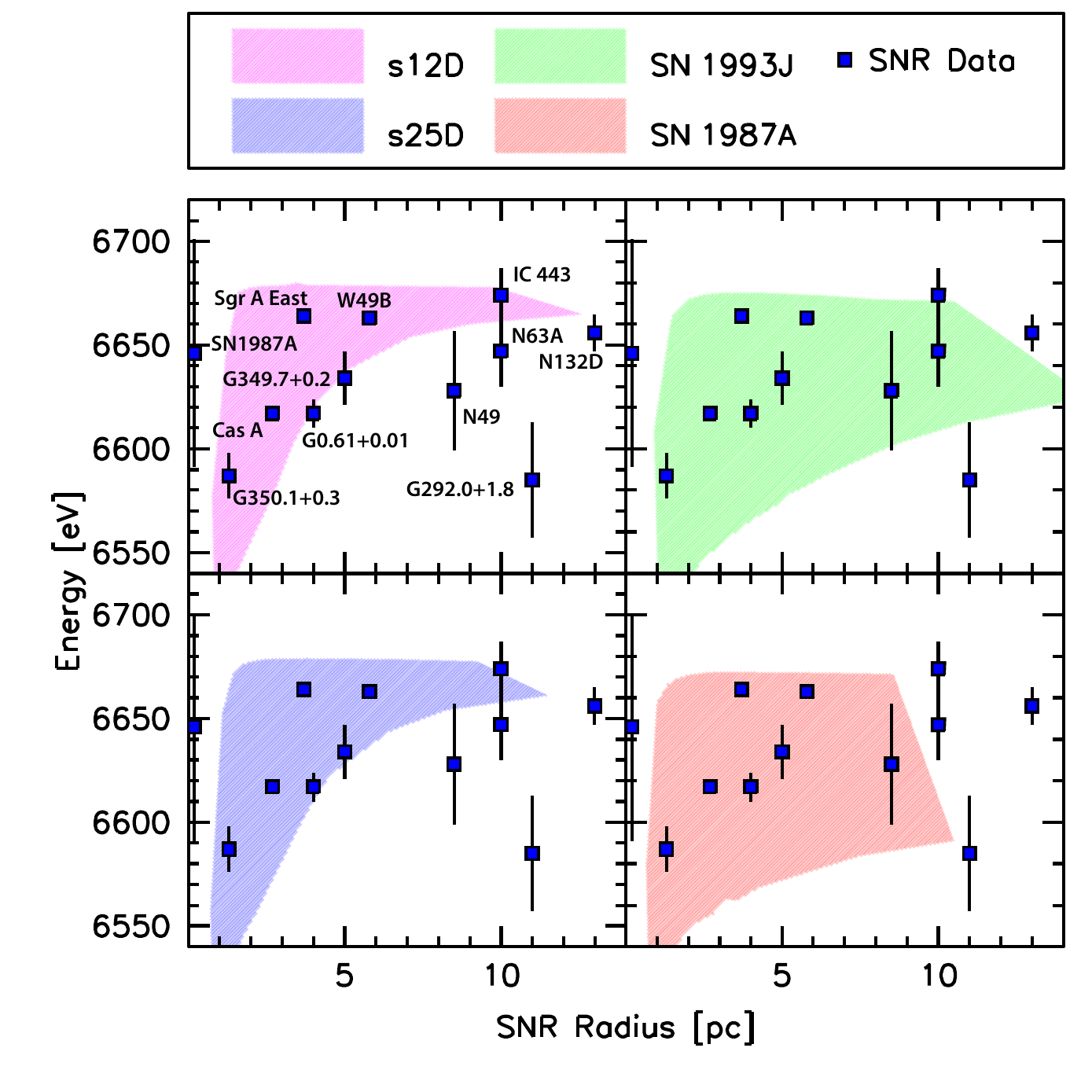}
\includegraphics[width=0.5\textwidth]{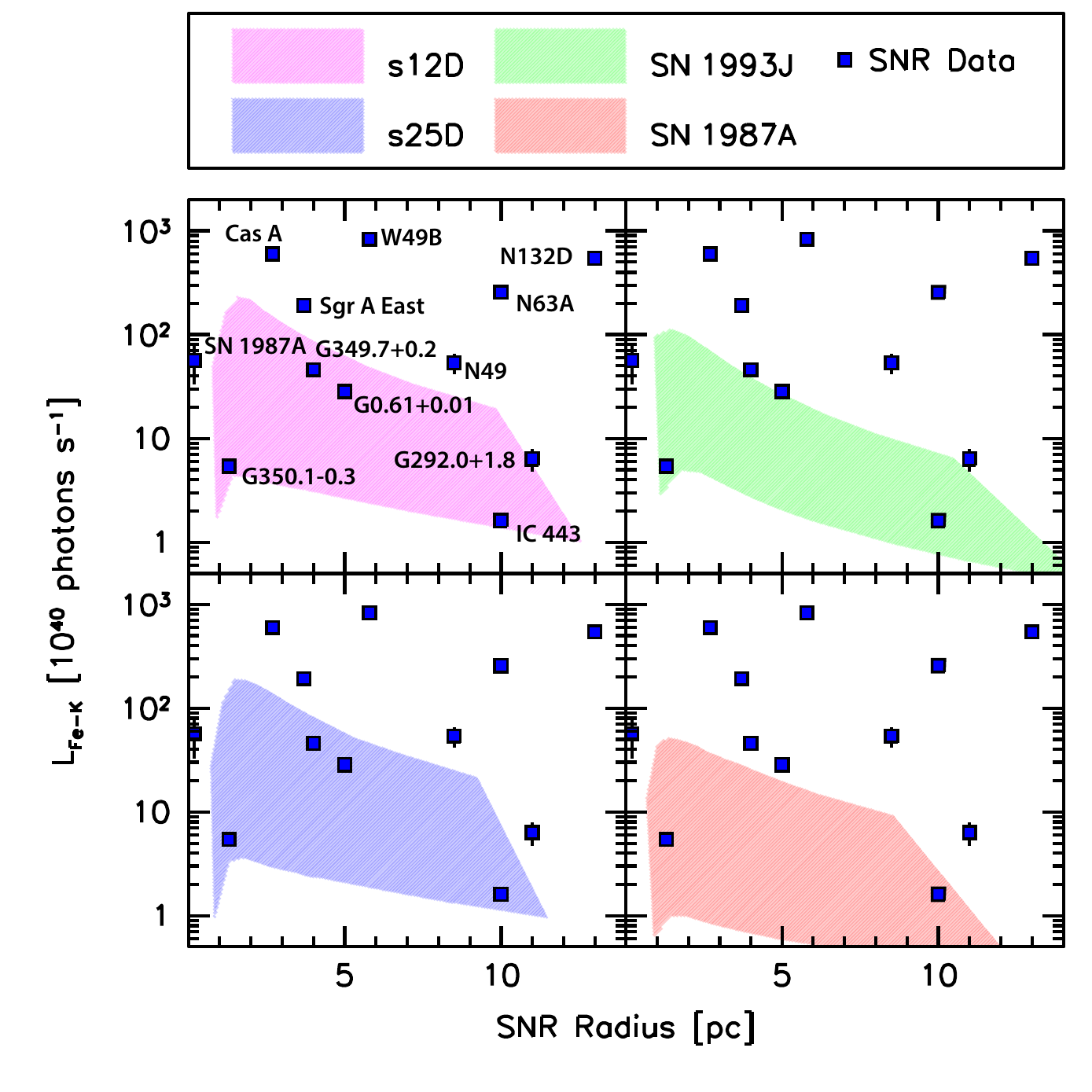}
\caption{{\it Left}: Fe-K line centroid energy as
a function of radius for core-collapse supernova remnant measurements
and CCSNe models listed in Table~\ref{tab:models}. {\it Right}:
Fe-K line luminosity as a function of radius for core-collapse supernova
remnant measurements and CCSNe models listed in Table~\ref{tab:models}.}
\label{fig:dyn_spec}
\end{figure}

As seen in Figure~\ref{fig:dyn_spec} (right), our modeled line luminosities
fall below the observed values for many remnants. Here
we argue that this may be associated with 
issues related to the very end stages
of the progenitor's evolution. While the observed SNR radii appear
consistent with the modeled radii, the luminosities differ by more than
an order of magnitude. So the question is whether one can increase the
luminosity without strongly impacting the radius. For core--collapse 
supernovae, the forward shock radius $R_b$ $\propto$ $\dot{M}^{-1/(n-s)}$
\citep{chevalier82}. Thus if one were to increase the mass--loss rate
prior to the supernova, we can estimate the change in blastwave radius
as a function of change in $\dot{M}$. First, from \citet{chevalier82}, 
the blastwave radius is:

\begin{equation}
R_b \propto \left[\frac{Ag^n}{q}\right]^{1/(n-s)}t^{\frac{n-3}{n-s}} \ ,
\end{equation}

\noindent
where $A$ is a constant, dependent upon the shape of the ejecta and
circumstellar environment, $n$ is the power law index that describes the
shape of the ejecta, and $s$ is the power law index that describes
the circumstellar environment. In this paper, $n=9$ and $s=2$. $g^n$ is
a constant, dependent upon the ejecta mass and explosion energy. $q$
is defined as $\dot{M}/4\pi v_w$. If the mass--loss rate is increased
by some amount $\chi$ prior to the supernova ($q^\prime = \chi q$), 
such that the 
blastwave radius is now $R_b^\prime \propto q^{(-1/(n-s))\prime}$, then
the fractional change in the radius, over the extent of time that the 
blastwave moves through the circumstellar shell is:

\begin{eqnarray}
\frac{R_b - R_b^{\prime}}{R_b} & = & \frac{q^{-1/(n-s)} - q^{(-1/(n-s))\prime}}{q^{-1/(n-s)}} \\
 & = & 1 - \chi^{-1/(n-s)} \ .
\end{eqnarray}

For large changes in the mass--loss rate, this effect can be quite
large. For instance, an order of magnitude increase in the mass--loss
rate results in a 30\% decrease in the blastwave radius. However, we
do not expect this increased mass--loss to persist for very long, so
the extent of the circumstellar medium occupied by the increased density
is small, essentially amounting to a thin circumstellar shell. 
After the blastwave breaks through this shell, it will accelerate
into the lower density slow RSG wind.

% For the core--collapse
% models, we consider ejecta with $\rho_{\mathrm{ej}}$ $\propto$ r$^{-n}$ 
% to some radius $R_c$ and then constant density inside of that radius,
% expanding into an isotropic wind with $\rho_{\mathrm w}$ = $qr^{-s}$. 
% At early times, the blastwave radius $R_b$ can be determined from
% \citet{chevalier82}:

% \begin{equation}
% R_c = [Ag^n/q]^{1/(n-s)}t^{(n-3)/(n-s)} \, .
% \end{equation}

% \noindent
% For our models, we assume $n=9$ and $s=2$, so from Table 1 of
% \citet{chevalier82}, $A$ = $0.096$ and $R_b/R_c$ = $1.250$. The 
% density profile in the power law region has

% \begin{equation}
% g^9 = \frac{25}{21\pi}\frac{E^2_{\mathrm{SN}}}{M_{\mathrm{ej}}} \, .
% \end{equation}

% So that

% \begin{equation}
% R_b = 0.77\left[\frac{E^{2}_{\mathrm{SN}}}{M_{\mathrm{ej}} q}\right]^{1/7}t^{6/7} \, .
% \end{equation}

% The constant $q$ = $\dot{M}/(4\pi v_{\mathrm{w}})$, which implies that
% $R_b$ $\propto$ $(\dot{M}/v_{\mathrm w})^{-1/7}$, or more generally,
% $\propto$ $(\dot{M}/v_{\mathrm w})^{1/(n-s)}$. For instance, if the 
% mass loss rate is increased by a factor of two, the early evolution
% of the blastwave radius only changes by less than 10\% (increasing the 
% mass loss rate alone causes a 10\% decrease in radius, but the 
% increased mass loss results in a lower ejecta mass so that more energy
% per gram is deposited into the supernova ejecta). Of course, this
% increased mass loss probably doesn't extend for a long time, so the 
% blastwave may expand into a lower density wind beyond this high mass loss
% phase.

In contrast, the X-ray luminosity L$_X$ $\sim$ n$^{2}_e \Lambda (T) V$, 
where $n_e$ is
the number of free electrons, $\Lambda(T)$ is the cooling function,
and $V$ is the emitting volume. Since $n_e$ $\propto$ $\dot{M}$, 
L$_X$ $\sim$ $\dot{M}^2$. A three-fold increase in the mass--loss
rate will result in an approximately order of magnitude increase in
L$_X$, an increase in collisional ionization, and thus an increase
in Fe-K line emission.
% Now consider the evolution of the X-ray luminosity and electron 
% temperature. L$_X$ $\sim$ n$^{2}_e \Lambda (T) V$, where $n_e$ is
% the number of free electrons, $\Lambda(T)$ is the cooling function,
% and $V$ is the emitting volume. Here, $n_e$ $\propto$ $\rho$ $\propto$
% $\dot{M}$. The cooling function is temperature dependent which, in turn,
% is dependent upon the shock velocity and density. Both the shock
% velocity and emitting volume ($V$) have a dependence upon the radius, 
% which in turn has a weak dependence upon the mass loss rate. Increasing
% the density will increase the luminosity without altering the
% temperature much. Also, since the density is higher, there will be
% more free particles for collisional ionization, giving the required
% Fe-K emission.
Thus, the problem posed by Figure~\ref{fig:dyn_spec} (right) can be 
qualitatively
addressed by altering the density of the circumstellar environment
close to the progenitor. This could be viewed as akin to higher
mass--loss just prior to the supernova. We note that this argument may break
down for shells that are sufficiently dense to cause significant 
radiative losses \citep{crowther07,pan13,smith14}.

In Figure~\ref{fig:dyn_spec} (left) we plot the modeled and observed
Fe-K line centroid versus SNR radii. As shown in Figure~\ref{fig:models},
the energy centroids for both the model and data are generally greater
than 6550 eV. In general, the observed and modeled line centroids as a 
function of energy are in agreement with one another. However, some
SNRs, G292.0+1.8 in particular, do not agree with any of the models. 
Interestingly, in the line luminosity versus radii plots, G292.0+1.8
also does not agree with the models. This may point to a more complicated
circumstellar environment around G292.0+1.8, evidence of which is seen
in X-ray observations \citep{park02,lee10}. However, in terms of the
models, we find that they generally predict Fe-K line centroids in excess
of 6550 eV for a broad range of radii. For the high mass--loss rate
models, the Fe-K line centroid is approaching collisional ionization
equilibrium -- increasing the mass--loss rate to affect a change in
the luminosity would not alter this conclusion.

\section{Conclusions}

We have presented an initial attempt to connect some properties of
core--collapse supernova, namely the composition and a parameterization of
the circumstellar environment, to some observable bulk properties 
of supernova remnants. We find that:

\begin{enumerate}

\item Using a model for the dynamical and spectral evolution of shocked
supernova ejecta and swept up material, we computed Fe-K line centroids 
and luminosities
for a broad range of parameters, including explosion energetics, 
ejecta mass, and circumstellar environment. We found that they distinctly
differ from those same properties computed from models for Type Ia 
supernovae.

\item We compared our models to measured Fe-K line properties and found
general agreement between the models and observations. For those SNRs that 
do not show good agreement with these simple models
we put forth possible reasons for the discrepancy.

\item We discuss the relationship between SNR radius and Fe-K luminosity,
and propose that higher mass--loss rates prior to the supernova may
be able to reconcile the discrepancy between the modeled and observed 
luminosities. The required higher mass--loss may arise either through
eruptive, episodic events prior to the SNe, such as in SN~2009ip, or
increased clumping in the wind during the later stages of evolution.
Both of these possibilities would result in a stellar wind profile
that differs from the $r^{-2}$ winds considered here.

\end{enumerate}

\begin{figure}
\includegraphics[width=0.5\textwidth]{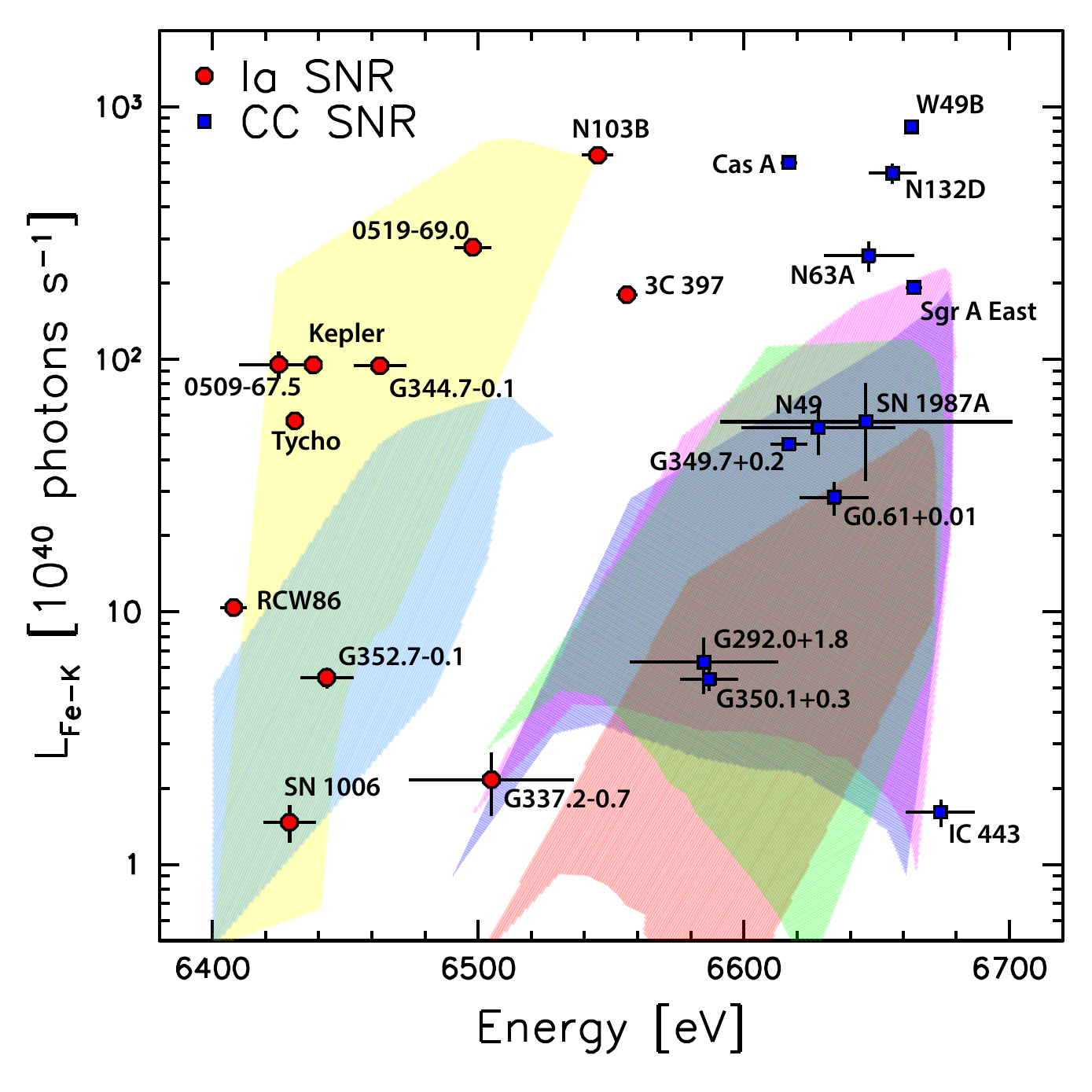}
\includegraphics[width=0.5\textwidth]{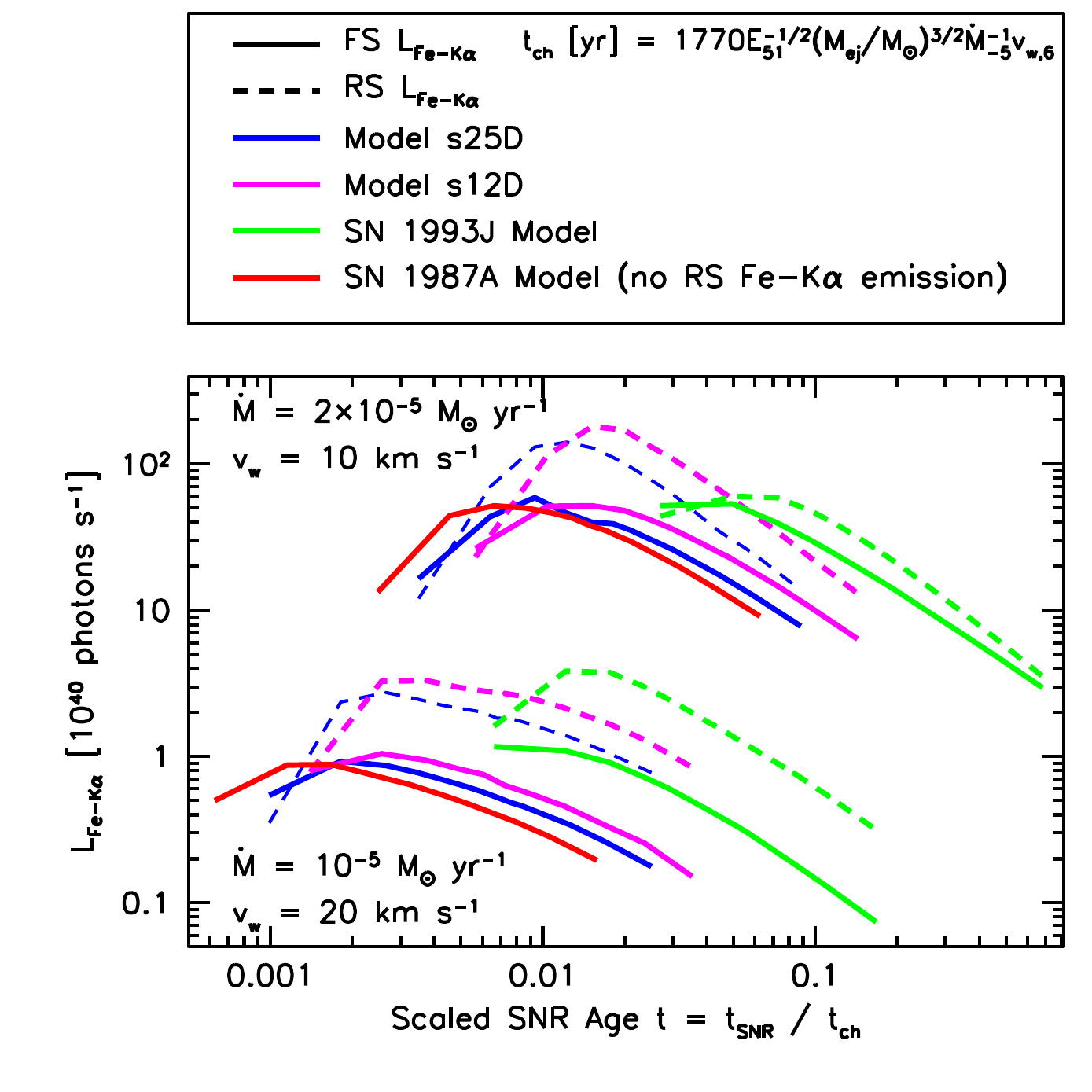}
\caption{{\it Left}: Fe-K line luminosity versus centroid energy for
Galactic and Magallenic Cloud Ia and core-collapse supernova remnants.
The transparent 
shaded regions correspond to the models listed in Table~\ref{tab:models}
with yellow corresponding to model DDTa, light blue to DDTg, red to 1987A,
green to 1993J, magenta to s12D and dark blue to s25D. The data are taken
from Table 1 of \citet{yamaguchi14}.{\it Right}: Fe-K luminosity as 
a function of scaled SNR age. The solid curves are the FS emission, while
the dashed curves are for the RS emission. The two sets of curves represent
the two sets of models listed in Table~\ref{tab:models}. The SN~1987A model
does not produce any appreciable Fe-K emission from the RS.}
\label{fig:models}
\end{figure}

For core--collapse supernovae, the range of ejecta mass, explosion energetics,
and circumstellar environment can be large. However, canonical values for
supernova parameters can reproduce many observations. Objects such as Cas A
and W49B require more detailed models, and any other SNR shown in
Figure~\ref{fig:models} may be scrutinized in more detail in order to 
pin down the exact parameter space. In terms of connecting supernovae 
(and their progenitor) models to supernova remnants, more detailed studies
of young SNRs such as SN~1993J, SN~1996c, and NGC~4449-1 are required 
in order to bridge the gap between truly young objects and Galactic
remnants. Additionally, studies of these objects will allow us to probe
the mass--loss history of the progenitor at times $\lesssim$ 10$^{5}$ years
before the explosion. Future concept X-ray missions with both
high imaging and spectral resolution, such as 
{\it Smart-X}\footnote{http://smart-x.cfa.harvard.edu} will foster
the detection and study of nearby extragalactic SNRs which can be added
to the sample discussed here.

Finally, understanding the mass--loss of supernova progenitors remains a 
challenge.
In this study, we simplified the circumstellar environment to be from a
a steady wind, though there is substantial evidence that steady line-driven
winds may not be prevalent. Understanding clumping in winds as well
as episodic mass--loss will be important, as these processes may enhance
the phenomena presented here. 

\acknowledgements

We thank R.~Fesen and D.~Milisavljevic for useful discussions during
the preparation of this manuscript. Additionally, we thank the anonymous
referee for the several useful suggestions which we have incorporated into
this paper.
D.~J.~P. acknowledges support for this work through the 
Smithsonian Institution's Competitive Grants for Science Program. D.~J.~P.
and P.~O.~S. also acknowledge support from NASA contract NAS8-03060. 
S.--H.~L. acknowledges support from Grants-in-Aid for Foreign JSPS Fellow
(No.~2503018). D.~C.~E. acknowledges support from NASA grant NNX11AE03G,
and S.~N. acknowledges support from the Japan Society for the Promotion
of Science (Nos. 23340069, 24.02022, 25.03786, and 25610056). 
A.~H.\ was supported by a Future Fellowship by the Australian 
Research Council (FT120100363).

\begin{deluxetable}{lcrrrcr}
\tablecolumns{7}
%%\tablewidth{0pc}
\tablecaption{Model Parameters}
\tablehead{
\colhead{Ejecta Model} & \colhead{E$_{SN}$} & \colhead{M$_{ej}$} 
& \colhead{$n_{amb}$\tablenotemark{a}} & \colhead{$v_{wind}$\tablenotemark{b}} 
& \colhead{$\dot{M}$\tablenotemark{b}} & \colhead{Ref.} \\
\colhead{} & \colhead{10$^{51}$ erg} & \colhead{M$_{\sun}$} & 
\colhead{cm$^{-3}$} & \colhead{km s$^{-1}$} & 
\colhead{10$^{-5}$ M$_{\sun}$ yr$^{-1}$} & \colhead{}}
\startdata
DDTa  & 1.27 & 1.38 & 0.1 -- 3.0 & \nodata & \nodata & \citet{badenes08} \\
DDTg  & 0.85 & 1.38 & 0.1 -- 3.0 & \nodata & \nodata & \citet{badenes08}\\
s12D  & 1.21 & 8.87 & \nodata & 10--20  & 1--2    & This work \\
s25D  & 1.21 & 12.2 & \nodata &  10--20  & 1--2   & This work \\
1987A & 1.10 & 14.7 & \nodata & 10--20  & 1--2    & \citet{saio88} \\
1993J & 2.00 & 2.92 & \nodata & 10--20  & 1--2    & \citet{nozawa10}
\enddata
\tablenotetext{a}{Type Ia models evolve in a constant density environment.}
\tablenotetext{b}{Core--collapse models evolve in a circumstellar environment
shaped by isotropic mass--loss.}
\label{tab:models}
\end{deluxetable}


\begin{thebibliography}

\bibitem[Badenes et al.(2003)]{badenes03} Badenes, C., Bravo, E., 
Borkowski, K.~J., \& Dom{\'{\i}}nguez, I.\ 2003, \apj, 593, 358 

\bibitem[Badenes et al.(2006)]{badenes06} Badenes, C., Borkowski, 
K.~J., Hughes, J.~P., Hwang, U., \& Bravo, E.\ 2006, \apj, 645, 1373 

\bibitem[Badenes et al.(2007)]{badenes07} Badenes, C., Hughes, 
J.~P., Bravo, E., \& Langer, N.\ 2007, \apj, 662, 472 

\bibitem[Badenes et al.(2008)]{badenes08} Badenes, C., Hughes, 
J.~P., Cassam-Chena{\"i}, G., \& Bravo, E.\ 2008, \apj, 680, 1149 


\bibitem[Bauer et al.(2008)]{bauer08} Bauer, F.~E., Dwarkadas, 
V.~V., Brandt, W.~N., et al.\ 2008, \apj, 688, 1210 

\bibitem[Blondin \& Lufkin(1993)]{blondin93} Blondin, J.~M., \& 
Lufkin, E.~A.\ 1993, \apjs, 88, 589 

\bibitem[Borkowski et al.(2007)]{borkowski07} Borkowski, K.~J., 
Hendrick, S.~P., \& Reynolds, S.~P.\ 2007, \apjl, 671, L45 

\bibitem[Chevalier(1982)]{chevalier82} Chevalier, R.~A.\ 1982, 
\apj, 258, 790 

\bibitem[Chevalier 
\& Emmering(1989)]{chevalier89} Chevalier, R.~A., \& Emmering, R.~T.\ 1989, \apjl, 342, L75 

\bibitem[Chevalier(2005)]{chevalier05} Chevalier, R.~A.\ 2005, 
\apj, 619, 839 

\bibitem[Chevalier et al.(2006)]{chevalier06} Chevalier, R.~A., 
Fransson, C., \& Nymark, T.~K.\ 2006, \apj, 641, 1029 

\bibitem[Chiotellis et 
al.(2012)]{chiotellis12} Chiotellis, A., Schure, K.~M., \& Vink, J.\ 2012, \aap, 537, A139 

\bibitem[Chugai(2009)]{chugai09} Chugai, N.~N.\ 2009, \mnras, 
400, 866 

\bibitem[Crowther(2007)]{crowther07} Crowther, P.~A.\ 2007, \araa, 45, 177 


\bibitem[Dwarkadas 
\& Chevalier(1998)]{dwarkadas98} Dwarkadas, V.~V., \& Chevalier, R.~A.\ 
1998, \apj, 497, 807 

\bibitem[Dwarkadas(2007)]{dwarkadas07} Dwarkadas, V.~V.\ 2007, 
\apj, 667, 226 

\bibitem[Dwarkadas(2011)]{dwarkadas11} Dwarkadas, V.~V.\ 2011, 
\mnras, 412, 1639 

\bibitem[Dwarkadas 
\& Gruszko(2012)]{dwarkadas12} Dwarkadas, V.~V., \& Gruszko, J.\ 2012, \mnras, 419, 1515 


\bibitem[Ellison et al.(2007)]{ellison07} Ellison, D.~C., 
Patnaude, D.~J., Slane, P., Blasi, P., \& Gabici, S.\ 2007, \apj, 661, 879 

\bibitem[Ellison et al.(2010)]{ellison10} Ellison, D.~C., 
Patnaude, D.~J., Slane, P., \& Raymond, J.\ 2010, \apj, 712, 287 

\bibitem[Ellison et al.(2012)]{ellison12} Ellison, D.~C., Slane, 
P., Patnaude, D.~J., \& Bykov, A.~M.\ 2012, \apj, 744, 39 

\bibitem[Fesen 
\& Matonick(1993)]{fesen93} Fesen, R.~A., \& Matonick, D.~M.\ 1993, \apj, 407, 110 

\bibitem[Filippenko(1997)]{filippenko97} Filippenko, A.~V.\ 1997, \araa, 35, 309 

\bibitem[Fransson et al.(1996)]{fransson96} Fransson, C., 
Lundqvist, P., \& Chevalier, R.~A.\ 1996, \apj, 461, 993 

\bibitem[Fransson et al.(2002)]{fransson02} Fransson, C., 
Chevalier, R.~A., Filippenko, A.~V., et al.\ 2002, \apj, 572, 350 

\bibitem[Gal-Yam et al.(2007)]{galyam07} Gal-Yam, A., Leonard, 
D.~C., Fox, D.~B., et al.\ 2007, \apj, 656, 372 

\bibitem[Hashimoto et 
al.(1989)]{hashimoto89} Hashimoto, M., Nomoto, K., \& Shigeyama, T.\ 1989, \aap, 210, L5 

\bibitem[Heger 
\& Woosley(2010)]{heger10} Heger, A., \& Woosley, S.~E.\ 2010, \apj, 724, 341 

\bibitem[Hughes(1987)]{hughes87} Hughes, J.~P.\ 1987, \apj, 314, 
103 

\bibitem[Hughes et al.(1995)]{hughes95} Hughes, J.~P., Hayashi, 
I., Helfand, D., et al.\ 1995, \apjl, 444, L81 

\bibitem[Hughes et al.(2000)]{hughes00} Hughes, J.~P., Rakowski, 
C.~E., Burrows, D.~N., \& Slane, P.~O.\ 2000, \apjl, 528, L109 

\bibitem[Hwang et al.(1993)]{hwang93} Hwang, U., Hughes, J.~P., 
Canizares, C.~R., \& Markert, T.~H.\ 1993, \apj, 414, 219 

\bibitem[Hwang 
\& Laming(2003)]{hwang03} Hwang, U., \& Laming, J.~M.\ 2003, \apj, 597, 362 

\bibitem[Hwang 
\& Laming(2012)]{hwang12} Hwang, U., \& Laming, J.~M.\ 2012, \apj, 746, 130 

\bibitem[Joggerst et al.(2009)]{joggerst09} Joggerst, C.~C., 
Woosley, S.~E., \& Heger, A.\ 2009, \apj, 693, 1780 

\bibitem[Koo \& McKee(1992)]{koo92} Koo, B.-C., 
\& McKee, C.~F.\ 1992, \apj, 388, 93 

\bibitem[Krause et al.(2008)]{krause08} Krause, O., Birkmann, 
S.~M., Usuda, T., et al.\ 2008, Science, 320, 1195 

\bibitem[Lee et al.(2010)]{lee10} Lee, J.-J., Park, S., 
Hughes, J.~P., et al.\ 2010, \apj, 711, 861 

\bibitem[Lee et al.(2014)]{lee14} Lee, S.-H., Patnaude, 
D.~J., Ellison, D.~C., Nagataki, S., \& Slane, P.~O.\ 2014, \apj, 791, 97 

\bibitem[Lodders et al.(2009)]{lodders09} Lodders, K., Palme, H., 
\& Gail, H.-P.\ 2009, Landolt B{\"o}rnstein, 44 

\bibitem[Lopez et al.(2013)]{lopez13a} Lopez, L.~A., 
Ramirez-Ruiz, E., Castro, D., \& Pearson, S.\ 2013, \apj, 764, 50 

\bibitem[Lopez et al.(2013)]{lopez13b} Lopez, L.~A., Pearson, 
S., Ramirez-Ruiz, E., et al.\ 2013, \apj, 777, 145 

\bibitem[Milisavljevic et al.(2012)]{milisavljevic12} Milisavljevic, 
D., Fesen, R.~A., Chevalier, R.~A., et al.\ 2012, \apj, 751, 25 

\bibitem[Moriya(2012)]{moriya12} Moriya, T.~J.\ 2012, \apjl, 
750, L13 

\bibitem[Nozawa et al.(2010)]{nozawa10} Nozawa, T., Kozasa, T., 
Tominaga, N., et al.\ 2010, \apj, 713, 356 

\bibitem[Ozawa et al.(2009)]{ozawa09} Ozawa, M., Koyama, K., 
Yamaguchi, H., Masai, K., \& Tamagawa, T.\ 2009, \apjl, 706, L71 

\bibitem[Pan et al.(2013)]{pan13} Pan, T., Patnaude, D., 
\& Loeb, A.\ 2013, \mnras, 433, 838 

\bibitem[Park et al.(2002)]{park02} Park, S., Roming, 
P.~W.~A., Hughes, J.~P., et al.\ 2002, \apjl, 564, L39 

\bibitem[Park et al.(2004)]{park04} Park, S., Hughes, J.~P., 
Slane, P.~O., et al.\ 2004, \apjl, 602, L33 

\bibitem[Patnaude et al.(2009)]{patnaude09} Patnaude, D.~J., 
Ellison, D.~C., \& Slane, P.\ 2009, \apj, 696, 1956 

\bibitem[Patnaude et al.(2010)]{patnaude10} Patnaude, D.~J., 
Slane, P., Raymond, J.~C., \& Ellison, D.~C.\ 2010, \apj, 725, 1476 

\bibitem[Patnaude et al.(2012)]{patnaude12} Patnaude, D.~J., 
Badenes, C., Park, S., \& Laming, J.~M.\ 2012, \apj, 756, 6 

\bibitem[Rauscher et al.(2002)]{rauscher02} Rauscher, T., Heger, 
A., Hoffman, R.~D., \& Woosley, S.~E.\ 2002, \apj, 576, 323 

\bibitem[Rest et al.(2008)]{rest08} Rest, A., Welch, D.~L., 
Suntzeff, N.~B., et al.\ 2008, \apjl, 681, L81 

\bibitem[Saio et al.(1988)]{saio88} Saio, H., Nomoto, K., 
\& Kato, M.\ 1988, \nat, 334, 508 

\bibitem[Salas et al.(2013)]{salas13} Salas, P., Bauer, F.~E., 
Stockdale, C., \& Prieto, J.~L.\ 2013, \mnras, 428, 1207 

\bibitem[Sana et al.(2012)]{sana12} Sana, H., de Mink, S.~E., 
de Koter, A., et al.\ 2012, Science, 337, 444 

\bibitem[Shigeyama 
\& Nomoto(1990)]{shigeyama90} Shigeyama, T., \& Nomoto, K.\ 1990, \apj, 360, 242 

\bibitem[Shigeyama et al.(1994)]{shigeyama94} Shigeyama, T., 
Suzuki, T., Kumagai, S., et al.\ 1994, \apj, 420, 341 

\bibitem[Smartt et al.(2002)]{smartt02} Smartt, S.~J., Gilmore, 
G.~F., Tout, C.~A., \& Hodgkin, S.~T.\ 2002, \apj, 565, 1089 

\bibitem[Smartt(2009)]{smartt09} Smartt, S.~J.\ 2009, \araa, 47, 63 

\bibitem[Smith et al.(2008)]{smith08} Smith, N., Chornock, R., 
Li, W., et al.\ 2008, \apj, 686, 467 

\bibitem[Smith et al.(2012)]{smith12} Smith, N., Mauerhan, 
J.~C., Silverman, J.~M., et al.\ 2012, \mnras, 426, 1905 

\bibitem[Smith(2014)]{smith14} Smith, N.\ 2014, \araa, 52, 487 

\bibitem[Soderberg et al.(2010)]{soderberg10} Soderberg, A.~M., 
Brunthaler, A., Nakar, E., Chevalier, R.~A., 
\& Bietenholz, M.~F.\ 2010, \apj, 725, 922 

\bibitem[Truelove 
\& McKee(1999)]{truelove99} Truelove, J.~K., \& McKee, C.~F.\ 
1999, \apjs, 120, 299 

\bibitem[Warren et al.(2003)]{warren03} Warren, J.~S., Hughes, 
J.~P., \& Slane, P.~O.\ 2003, \apj, 583, 260 

\bibitem[Weaver et al.(1978)]{weaver78} Weaver, T.~A., 
Zimmerman, G.~B., \& Woosley, S.~E.\ 1978, \apj, 225, 1021 

\bibitem[Woosley et al.(2002)]{woosley02} Woosley, S.~E., Heger, 
A., \& Weaver, T.~A.\ 2002, Reviews of Modern Physics, 74, 1015 

\bibitem[Woosley \& Heger(2007)]{woosley07} Woosley, S.~E., 
\& Heger, A.\ 2007, \physrep, 442, 269 

\bibitem[Yamaguchi et al.(2014)]{yamaguchi14} Yamaguchi, H., 
Badenes, C., Petre, R., et al.\ 2014, \apjl, 785, L27 

\end{thebibliography}
\end{document}